\documentclass[preprint,nofootinbib]{revtex4}
\usepackage{amssymb}
\usepackage{multirow}
\usepackage{array} 

\usepackage{hyperref}

\usepackage{amssymb}
\usepackage{times,fancyhdr}
\usepackage{amsmath}
\usepackage{amsfonts}
\usepackage{epstopdf}

\begin{document}

\newcommand{\be}{\begin{equation}}
\newcommand{\ee}{\end{equation}}
\def\bq{\begin{eqnarray}}
\def\eq{\end{eqnarray}}

\title{\bf  Relativistic potential of Weyl: a gateway to quantum physics in the presence of gravitation?}

\author{Ram Gopal Vishwakarma}

 \address{Unidad Acad$\acute{e}$mica de Matem$\acute{a}$ticas\\
 Universidad Aut$\acute{o}$noma de Zacatecas\\
 C.P. 98068, Zacatecas, ZAC, Mexico\\
Email: vishwa@uaz.edu.mx}

\begin{abstract}
The well-noted correspondence between gravitation and electrodynamics emphasizes the importance of the Lanczos tensor - the potential of the Weyl tensor - which is an inherent structural element of any metric theory of gravity formulated in a 4-dimensional pseudo Riemannian spacetime. However, this ingenious discovery has gone largely unnoticed. We elucidate this important quantity by deriving its expressions in some particularly chosen spacetimes and try to find out what it represents actually. We find out that the Lanczos potential tensor does not represent the potential of the gravitational field, as is ascertained by various evidences. Rather, it is enriched with various signatures of quantum character, which provides a novel insight into the heart of  a geometric embodiment of gravity.
\end{abstract}



\maketitle

\section{Introduction}

Einstein's revolutionary discovery of  the local equivalence of gravitation and inertia or the local cancellation
of the gravitational field by local inertial frames  - the (weak) equivalence principle - is one of the best-tested principles in the whole field of physics \cite{Will}. The principle is central to all metric theories of gravity, including Einstein's General Relativity (GR). As the gravitational field and the local inertial frames - both are characterized by the spacetime
metric, their equivalence helps to achieve the lofty scheme of the geometrization of gravitation in the geometrodynamical
structure of the pseudo-Riemannian spacetime.
Recognizing the metric field as the fundamental ingredient in a gravity theory, the Riemann tensor and the Weyl tensor are supposed to provide a deeper understanding of the geometric character of the theory. 
However, to further dissect the conceptual foundations of the Riemannian geometry, one must introduce additional
ideas that may in turn lead 
to a greater insight, thus helping in the modeling and the interpretation of the physical reality.

Due to the staggering amount of scientific and technological advancement made over the last few decades, our understanding of gravity has certainly improved quite a bit. Nevertheless, we cannot claim that all the consequences of the geometrization of gravity have already been fully explored. 
This is indicative of the fact that gravity has remained the most mysterious interaction  among the four known fundamental interactions.

In the following we address a fundamental feature of any metric theory of gravity - the gravitational analogue of  the electromagnetic potential - a comparatively unfamiliar and hitherto not seriously considered aspect of the theory. The study  sheds new light on the physical meaning of  the quantity indicating that it has a deeply ingrained quantum character, which may lead to a gateway to quantum physics in the presence of gravitation.

\section{Potential of the Weyl field}

In order to gain a deeper insight into the conceptual foundations of gravity, let us consider the well-known  correspondence\footnote{By this correspondence, we mean the fully covariant analogy  between gravity and electrodynamics emanating from the irreducible decomposition of the Weyl and the Maxwell tensors in the respective electric and magnetic parts. See, for example \cite{analogies}.} between gravity and electrodynamics, which has helped eminently in a better understanding of the gravity-problems again and again in the past.
In electrodynamics, a crucial ingredient of the electromagnetic field is its 4-potential $A_\mu$, ($\mu=0,1,2,3$)  whence emanates the Maxwell tensor $F_{\mu\nu}$:
\be
F_{\mu\nu} = A_{\mu;\nu} - A_{\nu;\mu}= A_{\mu,\nu} - A_{\nu,\mu},\label{eq:F}
\ee
which measures the strength of the field. Here the semicolon (comma) followed by an index denotes covariant (ordinary) derivative with respect to the corresponding variable. This demonstrates that it is the gauge field $A_\mu$, that is the fundamental field providing $F_{\mu\nu}$ as a derived concept through (\ref{eq:F}). Also,  in the Lagrangian for a free electromagnetic field, the basic field is $A_\mu$, and not $F_{\mu\nu}$. Variation of the Lagrangian with respect to $A_\mu$ gives the equations of motion.
Evidence for the direct detectability and physical importance of the potential field $A_\mu$  has already been given by the famous Aharonov-Bohm effect (discussed later).

What about the gravitational analogue of the electromagnetic 4-potential $A_\mu$?
As if in a direct answer to this question, an interesting feature of the 4-dimensional Riemannian geometry was discovered by Cornelius Lanczos during
the early part of the 1960’s \cite{Lanczos}:
That, there exists another classical characterization of the geometrical structure associated
with the Weyl conformal curvature tensor.
While analyzing the self-dual part of the Riemann tensor $R_{\mu\nu\sigma\rho}$ in four dimensions, Lanczos discovered a new tensor of rank three (now recognized as the Lanczos potential tensor $L_{\mu\nu\sigma}$) satisfying the symmetries
\be
L_{\mu\nu\sigma}=-L_{\nu\mu\sigma} ~~~~~~~~~~(a),~~~~~~~~~~ L_{\mu\nu\sigma}+L_{\nu\sigma\mu}+ L_{\sigma\mu\nu}=0 ~~~~~~~~~~(b) \label{eq:constraints}
\ee
and expressing linearly the Weyl tensor $C_{\mu\nu\sigma\rho}$ of the manifold in terms of the first covariant derivatives of the new tensor $L_{\mu\nu\sigma}$ through the generating equation
\be
C_{\mu\nu\sigma\rho} = L_{[\mu\nu][\sigma;\rho]} +  L_{[\sigma\rho][\mu;\nu]} -  {*L*}_{[\mu\nu][\sigma;\rho]} -  {*L*}_{[\sigma\rho][\mu;\nu]},\label{eq:Weyl-Lanczos1}
\ee
where the starred symbol denotes the dual operation defined by ${*N*}_{\alpha\beta\mu\nu}=\frac{1}{4}e_{\alpha\beta\rho\sigma}e_{\mu\nu\tau\delta} N^{\rho\sigma\tau\delta}$, with $e_{\mu\nu\sigma\rho}$ representing the Levi-Civita tensor and the square brackets [] denote antisymmetrization: for instance $X_{[\mu\nu]}\equiv \{X_{\mu\nu}-X_{\nu\mu}\}/2!$. 

Thus in the specific case of four dimensions and Lorentzian metric there indeed exists  a relativistic potential - the Lanczos potential tensor $L_{\mu\nu\sigma}$ - generating the Weyl tensor differentially,  in parallel to the electromagnetic gauge potential $A_\mu$ generating the field strength tensor $F_{\mu\nu}$.  This constitutes the Lanczos potential as a more fundamental geometrical object than the Weyl tensor. Let us recall that the Weyl tensor\footnote{The Riemann tensor is decomposed in terms of the Weyl and the Ricci tensors as 
\[
R_{\mu\nu\sigma\rho}=C_{\mu\nu\sigma\rho} - g_{\mu[\rho}R_{\sigma]\nu} - g_{\nu[\sigma}R_{\rho]\mu} - \frac{1}{3}Rg_{\mu[\sigma}~g_{\rho]\nu}.
\]
 The twenty degrees of freedom of the Riemann tensor are thus distributed equally among the Weyl and the Ricci tensors.}   $C_{\mu\nu\sigma\rho}$ is the gravitational analogue of the  electromagnetic Maxwell tensor $F_{\mu\nu}$. Both the tensors are trace-free. 

Later, a rigorous proof of existence was given for the Lanczos tensor generating the Weyl tensor of any 4-dimensional Riemannian manifold \cite{Bampi-Kaviglia}.  Interestingly, this potential tensor exists only for the Weyl tensor and not for the Riemann tensor in general \cite{Edgar}.
Thus the Lanczos potential emerges as a fundamental building block of a metric theory of gravity, deeply engraved in the respective Riemannian spacetime geometry, in the form of an inherent  structural element. 

Albeit its novelty and importance, this remarkable discovery is comparatively unfamiliar even now - some sixty years after Lanczos first introduced it -
and it has remained an obscure backwater to the mainstream relativists and cosmologists.
The main reason for this  connotative obscurity lies in the immense difficulty to calculate the Lanczos potential tensor for a given spacetime by integrating equation (\ref{eq:Weyl-Lanczos1}) directly, given the Weyl tensor. Although the Lanczos potentials have been investigated previously in several simple cases, there remains the lack of an algorithm that can allow one to obtain the Lanczos tensor unambiguously from lower-rank tensors for any given 4-dimensional pseudo-Riemannian spacetime in the most general case.

However, some progress in this direction has also been made, though heuristically.
Novello and Velloso have discovered by direct manipulation some algorithms, albeit ad-hoc, for Lanczos potential in terms of vector fields satisfying certain symmetries, which were then applied by them to calculate the potential tensor for Schwarzschild, Kasner and Godel spacetimes \cite{N-V}.
It appears that the Lanczos potential tensor can be obtained comparatively more easily by the use of the Newman-Penrose formalism. This formalism has been used to calculate the potential tensor for the Kerr and Petrov type N, III and O spacetimes in \cite{Bonilla1,Bonilla2}.

In order to exemplify the Lanczos potential, we first simplify the simple-looking and yet complex equation (\ref{eq:Weyl-Lanczos1}) by calculating the duals appearing in it. This yields
\bq\nonumber
C_{\mu\nu\sigma\rho}& = & L_{\mu\nu\sigma;\rho} +  L_{\sigma\rho\mu;\nu} -  L_{\mu\nu\rho;\sigma} -  L_{\sigma\rho\nu;\mu}
+g_{\nu\sigma}L_{(\mu\rho)}+g_{\mu\rho}L_{(\nu\sigma)}\\
& - & g_{\nu\rho}L_{(\mu\sigma)}-g_{\mu\sigma}L_{(\nu\rho)}
+\frac{2}{3}L^{\lambda\kappa}_{~~~\lambda;\kappa}(g_{\mu\sigma}g_{\nu\rho}-g_{\nu\sigma}g_{\mu\rho}),\label{eq:Weyl-Lanczos2}
\eq
where $L_{\mu\nu}\equiv L^{~~\kappa}_{\mu~~\nu;\kappa}-L^{~~\kappa}_{\mu~~\kappa;\nu}$ and 
the round brackets () denote symmetrization, i.e., $2X_{(\mu\nu)}\equiv X_{\mu\nu}+X_{\nu\mu}$.
Although the conditions  (\ref{eq:constraints}.a, b) are necessary and sufficient for $L_{\mu\nu\sigma}$ to generate $C_{\mu\nu\sigma\rho}$ through
 equation (\ref{eq:Weyl-Lanczos2}), Lanczos considered the following additional symmetries
\be
 L^{~~\kappa}_{\mu~~\kappa}=0 ~~~~~~~~~~(a),~~~~~~~~~~ L^{~~~\kappa}_{\mu\nu~~;\kappa}=0 ~~~~~~~~~~(b)\label{eq:gauge}
\ee
as two gauge conditions in order to reduce the number of degrees of freedom present in $L_{\mu\nu\sigma}$. 
He noticed that the Weyl tensor $C_{\mu\nu\sigma\rho}$ given by equation (\ref{eq:Weyl-Lanczos2}), remains invariant under the gauge transformation 
\be
\bar{L}_{\mu\nu\sigma} = L_{\mu\nu\sigma}+ g_{\nu\sigma}X_\mu - g_{\mu\sigma}X_\nu, \label{eq:gaugeU}
\ee
where $ X_\alpha$ is an arbitrary vector field. In order to fix this arbitrariness, he assumed the condition (\ref{eq:gauge}.a), which gives $ X_\alpha=0$. Whereas the condition (\ref{eq:gauge}.b) was adopted by him due to the reason that the divergence $ L^{~~~\kappa}_{\mu\nu~~;\kappa}$ does not participate in equation (\ref{eq:Weyl-Lanczos2}).
These however do not appear as compelling reasons to choose the Lanczos gauge (\ref{eq:gauge}) and one can adopt any other gauge by assigning any other (tensor) values to the tensors $ L^{~~\kappa}_{\mu~~\kappa}$ and $L^{~~~\kappa}_{\mu\nu~~;\kappa}$ suiting the considered problem.

\subsection{An elucidation of the Lanczos potential}

We try to exemplify the obscure theory of Lanczos potential in the following. For this purpose, the first example we consider is from the existing literature. We also calculate the potential tensor, for later use, in some other particularly chosen spacetimes.

\subsubsection{Schwarzschild Spacetime:}

Let us consider the Schwarzschild spacetime as the first example: 
\be
ds^2=\left(1-\frac{2m}{r}\right) dt^2-\frac{dr^2}{(1-2m/r)}-r^2d\theta^2-r^2\sin^2\theta ~d\phi^2.\label{eq:sch}
\ee
For simplicity, we have considered the geometric units with $G=1=c$. Novello and Velloso \cite{N-V} have shown that if a unit time-like vector field $V^\alpha\equiv dx^\alpha/ds$ tangential to the trajectory of an observer in a given spacetime is irrotational and shear-free, the Lanczos potential of the sapcetime is given by
\be
L_{\mu\nu\sigma}=V_{\mu;\kappa}V^\kappa V_\nu V_\sigma - V_{\nu;\kappa}V^\kappa V_\mu V_\sigma.\label{eq:NV}
\ee
By considering $V^\alpha=\left(\frac{1}{\sqrt{1-2m/r}},0,0,0\right)$, which comes out as irrotational and shear-free in the spacetime (\ref{eq:sch}), the formula (\ref{eq:NV}) calculates the corresponding Lancozs potential with only one non-vanishing (independent) component:
\be
L_{rtt}=-\frac{m}{r^2},\label{eq:Lan-sch1}
\ee
which though does not satisfy the gauge condition (\ref{eq:gauge}a) of trace-freeness, but it is divergence-free. A trace-free potential can be obtained by using relation (\ref{eq:gaugeU}), which allows to cancel the trace of the tensor by choosing $X_\mu=-L_{\mu~~\kappa}^{~~\kappa}/3$ giving 
\be\label{eq:Lan-sch}
\left.\begin{aligned}
&\bar{L}_{rtt}=-\frac{2m}{3r^2},\\
&\bar{L}_{r\theta\theta}=-\frac{m}{3(1-2m/r)},\\
&\bar{L}_{r\phi\phi}=-\frac{m\sin^2\theta}{3(1-2m/r)},
 \end{aligned}
 \right\}
\ee
which satisfy both Lanczos gauge conditions given in (\ref{eq:gauge}). This example illustrates that a spacetime can have different values of the potential tensor in different gauges.

\subsubsection{Kasner Spacetime:}

Next, we consider the Kasner spacetime, given by its line element in the form\footnote{This form of the Kasner line element is due to Narlikar and Karmarkar \cite{N-K}. The beauty of this form of the line element is that the coordinates $x,y,z$ appearing in it have the natural dimensions of length, unlike the standard form $ds^2= dt^2- t^{2p_1}dx^2- t^{2p_2}dy^2- t^{2p_3}dz^2$.}
\be
ds^2= dt^2- (1+nt)^{2p_1}dx^2- (1+nt)^{2p_2}dy^2
- (1+nt)^{2p_3}dz^2,\label{eq:kasner}
\ee
wherein $n$ is a dimensional parameter and $p_1$, $p_2$, $p_3$ dimensionless parameters satisfying
\[
p_1+p_2+p_3=1=p_1^2+p_2^2+p_3^2.
\]
Novello and Velloso have also shown that if the vector field $V^\alpha$ in a spacetime is irrotational and geodetic and satisfies the conditions ${*C}_{\mu\nu\lambda\rho} V^\nu V^\rho=0$ and 
$\sigma_{\mu\nu;\kappa}V^\kappa+\sigma_{\mu\nu} V^\kappa_{~~;\kappa}=0$, 
then the Lanczos tensor of the spacetime is given by
\[
L_{\mu\nu\lambda}=\frac{1}{3}\left(\sigma_{\mu\lambda}V_\nu - \sigma_{\nu\lambda}V_\mu\right),
\]
where $\sigma_{\mu\nu}$ represents shear in the vector field congruence $V^\alpha$.
Following Novello and Velloso, we consider $V^\alpha=\left(1,0,0,0\right)$, which satisfies all the prerequisites of the formula mentioned above for the line element (\ref{eq:kasner}). The non-vanishing (independent) components of $L_{\mu\nu\sigma}$ for this line element then come out as
\be\label{eq:Lan-kasner}
\left.\begin{aligned}
&L_{txx}=\frac{n}{9}(3p_1 -1)(1+nt)^{2p_1-1},\\
&L_{tyy}=\frac{n}{9}(3p_2 -1)(1+nt)^{2p_2-1},\\
&L_{tzz}=\frac{n}{9}(3p_3 -1)(1+nt)^{2p_3-1},
 \end{aligned}
 \right\}
\ee
which also satisfy, by chance, the Lanczos gauge conditions.

\subsubsection{Robertson-Walker Spacetimes:}

It may be curious to note that even when the Weyl tensor vanishes, its potential - the Lanczos tensor - can be non-vanishing. To illustrate this point, let us consider the Robertson-Walker (R-W) spacetime
\be
ds^2=dt^2-S^2(t)\left[\frac{dr^2}{1-kr^2}+r^2(d\theta^2+\sin^2\theta ~d\phi^2)\right],\label{eq:RW}
\ee
which is conformally flat and hence its Weyl tensor vanishes identically. 
It would be worth-while to mention that Hyoitiro Takeno \cite{Takeno} has derived differential equations for the Lanczos tensor  for  some particular cases of a spherically symmetric spacetime.
Taking inputs from this and using computational resources, we obtain the following as the Lanczos potential  for the spacetime (\ref{eq:RW}) with a single non-vanishing independent component
\be
L_{rtt}=\frac{a r}{\sqrt{1-kr^2}}, ~~~~~ a\equiv\text{an arbitrary constant}. \label{eq:L1-RW}
\ee
As this form is not trace-free, a trace-free form can be obtained, as mentioned earlier, by using equation (\ref{eq:gaugeU}), giving
\be\label{eq:L2-RW}
\left.\begin{aligned}
&\bar{L}_{rtt}=\frac{2a r}{3\sqrt{1-kr^2}},\\
&\bar{L}_{r\theta\theta}=\frac{a r^3S^2(t)}{3\sqrt{1-kr^2}},\\
&\bar{L}_{r\phi\phi}=\frac{a r^3S^2(t)}{3\sqrt{1-kr^2}}\sin^2\theta,
 \end{aligned}
 \right\}
\ee
 It may be noted that a spacetime can have more than one Lanczos potentials even in the same gauge. To exemplify this, we discover another set of 
$L_{\mu\nu\sigma}$ values satisfying the gauge condition (\ref{eq:gauge}.a) for the spacetime (\ref{eq:RW}):
\be\label{eq:L3-RW}
\left.\begin{aligned}
&\tilde{L}_{t \theta \phi}=b r^3\sin\theta,\\
&\tilde{L}_{t \phi \theta}=-b r^3\sin\theta,\\
&\tilde{L}_{\theta\phi t}=-2 br^3\sin\theta,
 \end{aligned}
 \right\}
\ee
where $b$ is an arbitrary constant. The presence of the arbitrary constants in the potentials given by (\ref{eq:L1-RW}-\ref{eq:L3-RW}) follows from the fact that given $L_{\mu\nu\sigma}$ as a solution of equation (\ref{eq:Weyl-Lanczos2}) with $C_{\mu\nu\sigma\rho}=0$, $a L_{\mu\nu\sigma}$ is also a solution with $a$ being an arbitrary constant. This happens because the Weyl tensor is linear in $L_{\mu\nu\sigma}$ in equation (\ref{eq:Weyl-Lanczos2}).
It is easy to check that a linear combination of the potentials given by (\ref{eq:L1-RW}-\ref{eq:L3-RW}) is also a potential of the spacetime (\ref{eq:RW}), owing to the same reason of the vanishing $C_{\mu\nu\sigma\rho}$ and the linearity of (\ref{eq:Weyl-Lanczos2}).

\subsubsection{R-W Spacetimes in Static Form:}

The Lanczos potentials of a given spacetime differ not only in different gauges, but also in different coordinates. To illustrate this, let us consider a static form of the spacetime (\ref{eq:RW}).
It is already known that this line element, which appears clearly dynamic in (\ref{eq:RW}), can be transformed to a static form in the case of a  constant Ricci scalar $R$ \cite{StaticFormRW}. 
For instance, for $k=1$ and $S=b\cosh(t/b)$, ~(where $b=$ constant), the line element can be transformed  to the static de-Sitter form  
\be
ds^2=\left(1-\frac{\rho^2}{b^2}\right) d\tau^2-\frac{d\rho^2}{(1-\rho^2/b^2)}-\rho^2(d\theta^2+\sin^2\theta ~d\phi^2)\label{eq:staticCRW}
\ee
by use of the transformations $\rho=b~r\cosh(t/b), ~ \tanh(\tau/b)=(1-r^2)^{-1/2}\tanh(t/b)$. We can now apply the formula (\ref{eq:NV}) of Novello-Velloso to calculate the Lanczos potential of (\ref{eq:staticCRW}).
By considering $V^\alpha=\left(\frac{1}{\sqrt{1-\rho^2/b^2}},0,0,0\right)$, which fulfills the prerequisites of the formula, the non-vanishing independent components of $L_{\mu\nu\sigma}$ yield
\be
L_{\rho\tau\tau}=a\frac{\rho}{b^2},\label{eq:Lan-CRW1}
\ee
in the gauge (\ref{eq:gauge}.b) and
\be\label{eq:Lan-CRW}
\left.\begin{aligned}
&\bar{L}_{\rho\tau\tau}=\frac{2a\rho}{3b^2},\\
&\bar{L}_{\rho\theta\theta}=-\frac{a\rho^3}{3(\rho^2-b^2)},\\
&\bar{L}_{\rho\phi\phi}=-\frac{a\rho^3\sin^2\theta}{3(\rho^2-b^2)},
 \end{aligned}
 \right\}
\ee
in the Lanczos gauge (\ref{eq:gauge}.a, b). Here $a$ is an arbitrary constant.

Similarly, in the case $k=-1$, the R-W line element (\ref{eq:RW})
can be transformed, for $S=b\sin(t/b)$, to its static form
\be
ds^2=\left(1+\frac{\rho^2}{b^2}\right) d\tau^2-\frac{d\rho^2}{(1+\rho^2/b^2)}-\rho^2(d\theta^2+\sin^2\theta ~d\phi^2)\label{eq:staticORW}
\ee
by use of the transformations $\rho=b~r\sin(t/b), ~ \tan(\tau/b)=\sqrt{(1+r^2)}\tan(t/b)$.
By considering $V^\alpha=\left(\frac{1}{\sqrt{1+\rho^2/b^2}},0,0,0\right)$, the formula (\ref{eq:NV}) gives the non-vanishing independent components of the Lanczos potential as
\be
L_{\rho\tau\tau}=-a\frac{\rho}{b^2},\label{eq:Lan-ORW1}
\ee
in the gauge (\ref{eq:gauge}.b) and
\be\label{eq:Lan-ORW}
\left.\begin{aligned}
&\bar{L}_{\rho\tau\tau}=-\frac{2a\rho}{3b^2},\\
&\bar{L}_{\rho\theta\theta}=-\frac{a\rho^3}{3(\rho^2+b^2)},\\
&\bar{L}_{\rho\phi\phi}=-\frac{a\rho^3\sin^2\theta}{3(\rho^2+b^2)},
 \end{aligned}
 \right\}
\ee
in the Lanczos gauge.

\subsection{A misunderstanding creped into the literature}

There appears a widespread misunderstanding\footnote{A similar issue related with the Lanczos tensor has been dealt with in \cite{JMP}.} in the literature regarding the degrees of freedom/number of independent components of $L_{\mu\nu\sigma}$. It is obvious that the antisymmetry in the first two indices of $L_{\mu\nu\sigma}$ represented by condition (\ref{eq:constraints}.a) leaves the number of independent components of $L_{\mu\nu\sigma}$ as 24. Further, the cyclic symmetry (\ref{eq:constraints}.b), which is equivalent to  ${*L}^{~~\kappa}_{\mu~~\kappa}=0$, gives 4 independent equations hence reducing the number of independent components of $L_{\mu\nu\sigma}$ to 20. 
Lanczos tried to further reduce the number of degrees of freedom of $L_{\mu\nu\sigma}$ in order to match that of the Weyl, which is 10. 
As has been explained earlier, he assumed the condition (\ref{eq:gauge}.a), to fix the arbitrariness in $L_{\mu\nu\sigma}$ brought about through (\ref{eq:gaugeU}). This algebraic gauge condition (\ref{eq:gauge}.a) provides 4 equations and reduces the number of independent components of $L_{\mu\nu\sigma}$ to 16. Up to this point everything goes fine.

However, claiming that the differential gauge condition (\ref{eq:gauge}.b), which provides 6 equations, reduces the number of independent components of $L_{\mu\nu\sigma}$ to 10 and thus matches the number of independent components of Weyl  thereby providing a unique $L_{\mu\nu\sigma}$ in a given spacetime (which is largely\footnote{Though  a few authors, for example \cite{Bonilla1}, have mentioned the non-uniqueness in the value of $L_{\mu\nu\sigma}$.} posited in the existing literature), does not seem correct. In fact, the differential gauge condition is not effective in reducing the number of independent components of the tensor. 
The reason why this is so is the following.

The differential gauge condition (\ref{eq:gauge}.b), unlike its algebraic counterpart (\ref{eq:gauge}.a),  does not in general supply 6 clean algebraic equations in the components of $L_{\mu\nu\sigma}$ only.
Rather the integration of the partial differential equations resulting from the condition (\ref{eq:gauge}.b), also produces arbitrary functions which provide `handles' one can randomly adjust.  In order to fix this arbitrariness, one needs additional conditions/assumptions. Thus the condition (\ref{eq:gauge}.b) alone, taken together with (\ref{eq:constraints}.a, b) and (\ref{eq:gauge}.a), cannot supply a unique value of $L_{\mu\nu\sigma}$ in a given spacetime.
As we shall soon see in an example, an abundant degeneracy in the value of $L_{\mu\nu\sigma}$  appears even after the gauge conditions (\ref{eq:gauge}.a, b) are applied.

Let us recall that it is only the algebraic symmetries of the  Riemann (Weyl) tensor which determine its 20 (10) independent components, and the differential symmetries (Bianchi identities) do {\it  not} contribute to it. Similarly, the Weyl tensor admits a divergence-free condition $C^\mu_{~~\nu\sigma\rho;\mu}=0$ in the Ricci-flat spacetimes ($R_{\mu\nu}=0$), but this does {\it not} further reduce its 10 independent components.

Interestingly, the above-mentioned arbitrariness in $L_{\mu\nu\sigma}$ can be attributed to another agent discovered by Takeno in \cite{Takeno}. 
He noticed that given a Lanczos potential $L_{\mu\nu\sigma}$ of a particular spacetime, the quantity 
\be
\bar{L}_{\mu\nu\sigma}=L_{\mu\nu\sigma}+A_{\mu\nu\sigma}\label{eq:barL}
\ee
 is again a Lanczos potential of that spacetime if  the tensor $A_{\mu\nu\sigma}$ (termed as `s-tensor' by Takeno) satisfies 
\be\label{eq:A}
\left.\begin{aligned}
A_{\mu\nu\sigma}=-A_{\nu\mu\sigma},~~~~~~~~~~~~~~~~ A_{\mu\nu\sigma}+A_{\nu\sigma\mu}+ A_{\sigma\mu\nu}=0,\\
A_{[\mu\nu][\sigma;\rho]} +  A_{[\sigma\rho][\mu;\nu]} -  {*A*}_{[\mu\nu][\sigma;\rho]} -  {*A*}_{[\sigma\rho][\mu;\nu]}=0.
 \end{aligned}
 \right\}
\ee
Thus the tensor $A_{\mu\nu\sigma}$ appears as an auxiliary potential tensor for the considered spacetime, as is clear from a comparison of equation (\ref{eq:A}) with (\ref{eq:constraints}) and (\ref{eq:Weyl-Lanczos1}). Obviously the gauge condition (\ref{eq:gauge}.b) of divergence-freeness, which implies that $A^{~~~\kappa}_{\mu\nu~~;\kappa}=0$,  does not fix  $A_{\mu\nu\sigma}$ uniquely owing to the same reason as mentioned above and hence causes degeneracy in $L_{\mu\nu\sigma}$. This is clear from the following examples, which show the existence of more than single values of $A_{\mu\nu\sigma}$ and hence $L_{\mu\nu\sigma}$ in the Schwarzschild spacetime, all values satisfying the Lanczos gauge conditions (\ref{eq:gauge}.a, b). 

As has been mentioned above, Takeno \cite{Takeno} has also derived differential equations for the tensor $A_{\mu\nu\sigma}$ in Lanczos gauge for some particular cases of a spherically symmetric spacetime. It can be solved for the Schwarzschild line element (\ref{eq:sch}) yielding
\be\label{eq:A1}
\left.\begin{aligned}
&A_{trr}=\frac{2}{r^3(1-2m/r)},\\
&A_{t\theta\theta}=-\frac{1}{r},\\
&A_{t\phi\phi}=-\frac{\sin^2\theta}{r},
 \end{aligned}
 \right\}
\ee
as the non-vanishing independent components of the auxiliary potential tensor for this spacetime in Lanczos gauge (i.e., they satisfy $A^{~~\kappa}_{\mu~~\kappa}=0$ and $A^{~~~\kappa}_{\mu\nu~~;\kappa}=0$ in addition to (\ref{eq:A})). 
We find another solution for the line element (\ref{eq:sch}) as
\be\label{eq:A2}
\left.\begin{aligned}
&A_{rtt}=2r,\\
&A_{r\theta\theta}=\frac{r^3}{1-2m/r},\\
&A_{r\phi\phi}=\frac{r^3\sin^2\theta}{1-2m/r},
 \end{aligned}
 \right\}
\ee
in the same Lanczos gauge.
Interestingly, the potential tensor (\ref{eq:L3-RW}), calculated for the spacetime \ref{eq:RW}, also forms yet another auxiliary potential for the  Schwarzschild spacetime (\ref{eq:sch}) in the Lanczos gauge\footnote{It may be interesting to note that the solution given by (\ref{eq:L3-RW}) appears as an auxiliary potential tensor for many (all?) spherically symmetric spacetimes, for instance the spacetimes (\ref{eq:sch},\ref{eq:RW},\ref{eq:Mink}). Perhaps this has some deeper physical meaning, unveiling thereof requires further study. Let us note that an auxiliary potential becomes a proper potential of a spacetime whose Weyl tensor vanishes, as is clear from a comparison of equation (\ref{eq:A}) with (\ref{eq:constraints}) and (\ref{eq:Weyl-Lanczos1}).}.
Clearly the conditions (\ref{eq:A}) imply that $a A_{\mu\nu\sigma}$ is also an auxiliary potential of the considered spacetime for an arbitrary constant $a$, given that $A_{\mu\nu\sigma}$ is an auxiliary potential tensor. Similarly, a linear combination of two or more auxiliary potentials of a given spacetime is again an auxiliary potential. 
Thus, these examples of the auxiliary potential of the Schwarzschild spacetime, taken together with its Lanczos potential given by (\ref{eq:Lan-sch}),  provide ample evidence for the degeneracy in the values of the tensor $A_{\mu\nu\sigma}$ and hence the degeneracy in $L_{\mu\nu\sigma}$ (through (\ref{eq:barL})) even after the Lanczos gauge conditions are applied. 

This illustrates that the cause of the degeneracy in $L_{\mu\nu\sigma}$ is not an issue with the gauge but it arises due to the redundant degrees of freedom of the tensor.
Similar situation appears in electrodynamics where different electromagnetic four-potentials correspond to the same electromagnetic field, depending upon the choice of gauge, or even in the same gauge \cite{Reiss}.

\section{On the physical meaning of the Lanczos tensor}

The great formal similitude between gravitation and electrodynamics indicates that the Lanczos potential must be imbued with interesting physical properties. Nevertheless, this terrain is largely unexplored and Lanczos's ingeneous discovery has remained more or less a mathematical curiosity. This is also one of the reasons of the virtual obscurity of Lanczos's theory. Here we attempt to explore what the Lanczos potential tensor $L_{\mu\nu\sigma}$ may represent physically. In this endeavor we first show what $L_{\mu\nu\sigma}$ does not represent.

\subsection{$L_{\mu\nu\sigma}$ does not represent the potential of the gravitational field}

Since the Lanczos tensor appears as the potential  to the Weyl tensor and since the latter is linked with the gravitational field, one may naturally expect the Lanczos tensor to represent the relativistic formulation of the gravitational potential. However, if this is so, the tensor is expected to reduce to the Newtonian potential  in a weak gravitational field.
As the Newtonian theory of gravitation provides excellent approximations under a wide range of
astrophysical cases, the first crucial test of any theory of gravitation is that it reduces to the Newtonian
gravitation in the limit of a weak gravitational field where the velocities are small compared with the speed of light. This requirement however does not seem to be fulfilled by $L_{\mu\nu\sigma}$, as we shall see in the following.

Let us consider a static point mass $m$ placed at the origin of a centrally symmetric coordinate system $r,\theta,\phi$.
In the Newtonian theory of gravity, the gravitational field produced by the mass at a point $r$ is represented in terms of the gravitational potential $\Phi(r)=-m/r$ at that point.
 In a relativistic theory of gravitation, for example GR, the gravitational field of the mass is well-described by the Schwarzschild line element (\ref{eq:sch}). In the case of a weak field, when the spacetime line element differs minutely from the Minkowskian metric $\eta_{\mu\nu}$ given by \ref{eq:Mink}), i.e.
\be
g_{\mu\nu}= \eta_{\mu\nu}+ h_{\mu\nu},~~~~~~{\rm where} ~~~~|h_{\mu\nu}|<<1,\label{eq:weak}
\ee
the line element (\ref{eq:sch}) reduces to
\be
ds^2=\eta_{\mu\nu}dx^\mu dx^\nu -\frac{2m}{r}(dt^2+dr^2),\label{eq:schW}
\ee
in the first order of approximation [$(h_{\mu\nu})^2<<h_{\mu\nu}$], hence giving the only non-vanishing components of $h_{\mu\nu}$ as $h_{tt}=h_{rr}=-2m/r$.
It has been shown \cite{Lanczos} that the Lanczos tensor, in the case of (\ref{eq:weak}), can be written in terms of the metric tensor as
\be
L_{\mu\nu\sigma}=\frac{1}{4}\left(h_{\mu\sigma,\nu} - h_{\nu\sigma,\mu} +\frac{1}{6} h_{,\mu}\eta_{\nu\sigma} - \frac{1}{6} h_{,\nu}\eta_{\mu\sigma}\right),   ~~~~ ~~~h\equiv  h_{\mu\nu}\eta^{\mu\nu}.\label{eq:weakL}
\ee 
 in the first order of approximation.
 For the line element (\ref{eq:schW}), this definition then provides the only non-vanishing independent component of $L_{\mu\nu\sigma}$ as
\[
L_{rtt}=-\frac{m}{2r^2},
\]
which though does not match with the Newtonian value of the gravitational potential $\Phi(r)=-m/r$ (This was expected since it is the metric tensor, and not its derivatives [as in (\ref{eq:weakL})], that gives the Newtonian potential in a weak field \cite{ABS}.).
Thus $L_{\mu\nu\sigma}$ does not seem to represent the potential of the gravitational field. This is corroborated by another evidence - inconsistencies in the construction of an energy-momentum tensor of the gravitational field from $L_{\mu\nu\sigma}$.

\subsection{An energy-momentum tensor from the Lanczos tensor}

Like the energy-momentum of matter, the energy-momentum of the gravitational field itself gravitates.
Hence, there have been many dedicated efforts to construct the energy-momentum tensor of the gravitational field. However, the quantities which are generally arrived at for this purpose are various energy-momentum pseudotensors, which are known to be unsatisfactory. An important breakthrough in this direction is the formulation of a completely symmetric and trace-free tensor of rank 4 known as the Bel-Robinson tensor 
\be
\overset{\rm BR}{T}_{\alpha\beta\gamma\delta} = C^{~~\sigma\rho}_{\alpha~~~\gamma}~ C_{\beta\sigma\rho\delta} + *C^{~~\sigma\rho}_{\alpha~~~\gamma}~ {*C}_{\beta\sigma\rho\delta},\label{eq:B-R}
\ee
which has been derived from the Weyl tensor in analogy to the energy-momentum tensor of the electromagnetic field \cite{B} (for another interpretation of the tensor, see \cite{IJGMMP}). However, the Bel-Robinson tensor has the wrong dimensions: dimensions of the energy density squared. Incidentally, a tensor constructed along the lines of (\ref{eq:B-R}) from the Lanczos tensor, has the correct dimensions (dimensions of the energy density), which was introduced in \cite{Roberts} as a possible candidate of the energy-momentum tensor of the gravitational field. A tensor of rank 4 formulated along the lines of (\ref{eq:B-R}) out of $L_{\mu\nu\sigma}$ is 
\be
T_{\alpha\beta\gamma\delta} = L^{~~\sigma}_{\alpha~~~\gamma}~ L_{\beta\sigma\delta} + *L^{~~\sigma}_{\alpha~~~\gamma}~ {*L}_{\beta\sigma\delta},\label{eq:BRL4}
\ee
though it is not symmetric in all pair of indices, contrary to what one expects from an energy-momentum tensor. A symmetric tensor can be obtained from (\ref{eq:BRL4}) by contracting over the last pair of indices of $T_{\alpha\beta\gamma\delta}$, giving
\be
T_{\alpha\beta} = L^{~~\sigma\rho}_{\alpha}~ L_{\beta\sigma\rho} + *L^{~~\sigma\rho}_{\alpha}~ {*L}_{\beta\sigma\rho},\label{eq:BRL}
\ee
which is symmetric and also trace-free (even if $L_{\mu\nu\sigma}$ does not satisfy the trace-free condition) like the Bel-Robinson tensor. (However, $T_{\alpha\beta}$ is not divergence-free in general.)
It is expected that a completely time-like component of the energy-momentum tensor relative to any observer must be positive definite - a desirable property for any candidate of the energy density of the gravitational field.
Nevertheless, the tensor $T_{\alpha\beta}$ given by (\ref{eq:BRL}) does not seem to fulfill this requirement, as we shall see in the following.

By defining the expected energy density $E$ of the gravitational filed measured by a stationary observer described by the timelike unit
vector field $V^\alpha$ by
\[
E=T_{\alpha\beta}~V^\alpha V^\beta,
\]
we calculate $E$ for different spacetimes considered earlier. For the Schwarzschild spacetime with $V^\alpha=\left(\frac{1}{\sqrt{1-2m/r}},0,0,0\right)$, the values of the Lanczos tensor given by (\ref{eq:Lan-sch1}) and (\ref{eq:Lan-sch}) generate the values of $E$  respectively as 
\[
E=- \frac{m^2}{r^4(1-2m/r)}, ~~~~~ E=- \frac{2m^2}{9r^4(1-2m/r)},
\]
which are negative-definite for $r>2m$. Similarly, for the  R-W spacetimes with $V^\alpha=\left(1,0,0,0\right)$, its Lanczos potentials  given by (\ref{eq:L1-RW}), (\ref{eq:L2-RW})  and (\ref{eq:L3-RW}) generate respectively 
\[
E  =  -\frac{a^2r^2}{S^2},~~~~E=-\frac{2a^2r^2}{9S^2},~~~~ E=-\frac{b^2r^2}{S^4},
\]
which are negative-definite for all values of $r$, $S(t)$ and $a,b$. 
For the open static R-W spacetime (\ref{eq:staticORW}) with $V^\alpha=\left(\frac{1}{\sqrt{1+\rho^2/b^2}},0,0,0\right)$, its Lanczos potentials (\ref{eq:Lan-ORW1}) and (\ref{eq:Lan-ORW}) generate respectively 
\[
E=-\frac{a^2\rho^2}{b^2(b^2+\rho^2)}, ~~~~     E = -\frac{2a^2\rho^2}{9b^2(b^2+\rho^2)},
\]
which are negative-definite for all values of $\rho$ and the constants $a,b$. Similarly, for the closed static R-W spacetime  (\ref{eq:staticCRW}) with $V^\alpha=\left(\frac{1}{\sqrt{1-\rho^2/b^2}},0,0,0\right)$, its Lanczos potentials (\ref{eq:Lan-CRW1}) and (\ref{eq:Lan-CRW}) generate respectively
\[
E=-\frac{a^2\rho^2}{b^2(b^2-\rho^2)}, ~~~~ E=-\frac{2a^2\rho^2}{9b^2(b^2-\rho^2)},
\]
which are negative for $\rho<b$ with any value of the constants $a,b$.
However, $E$ is not negative for all spacetimes. Rather its sign seems arbitrary. For instance, it has a positive-definite value
for the Kasner spacetime. For this spacetime with $V^\alpha=\left(1,0,0,0\right)$, its Lanczos tensor given by (\ref{eq:Lan-kasner}) provide
\[
E=\frac{2n^2}{27(1+nt)^2},
\]
which is positive-definite for all values of $n$ and $t$. 

There is another, and even more important, requirement which is expected to be satisfied by a genuine energy-momentum tensor. It can be described by
\[
E=0  \Leftrightarrow  T_{\alpha\beta}=0  \Leftrightarrow  L_{\alpha\beta\gamma}=0.
\]
However, the definition (\ref{eq:BRL}) does not seem to satisfy this fundamental property in general. This can be checked by generalizing the Lanczos potential (\ref{eq:L2-RW}) of the R-W spacetime by
\be\label{eq:Ln-RW}
\left.\begin{aligned}
& L_{rtt}=\frac{a_1 r}{\sqrt{1-kr^2}},\\
& L_{r\theta\theta}=\frac{a_2 r^3S^2(t)}{\sqrt{1-kr^2}},\\
& L_{r\phi\phi}=\frac{a_2 r^3S^2(t)}{\sqrt{1-kr^2}}\sin^2\theta,
 \end{aligned}
 \right\}
\ee
in another gauge. It can be easily checked that this value of $ L_{\alpha\beta\gamma}$ indeed satisfies the generating equation (\ref{eq:Weyl-Lanczos2}) for arbitrary values of the constants $a_1,a_2$ for the line element (\ref{eq:RW}). Interestingly, the potential (\ref{eq:Ln-RW}), for the unit timelike vector $V^\alpha=\left(1,0,0,0\right)$, gives
\[
E=-\frac{(a_1^2-2a_2^2)r^2}{S^2},
\]
which vanishes by choosing the constants $a_1,a_2$ through $a_1=\sqrt{2}a_2$, though $T_{\alpha\beta}$ and $ L_{\alpha\beta\gamma}$ do not vanish for this choice of the constants in general.

These undesired results simply indicate that the Lanczos tensor $L_{\mu\nu\sigma}$ cannot be the potential of the gravitational field. That is what these mathematical results cry out.

\section{Quantum effects through the Lanczos potential}

If the Lanczos tensor does not represent the potential of the gravitational field, what else does it represent then, in the presence of gravity ascribed to the spacetime curvature? The electromagnetic analogy again helps to seek the answer to this question.

In classical electrodynamics, the electromagnetic 4-potential is generally considered no more than a mathematical tool for solving the Maxwell equations. However, it has an unavoidable role at quantum level,  and convey physical information beyond what is supplied by fields alone that are derived from it.
Aharonov and Bohm \cite{A-B} first pointed out the reality and importance of the potentials in quantum realms (Aharonov-Bohm effect), which was confirmed experimentally by Chambers \cite{Chambers}.

The perfect analogy between electrodynamics and gravitation, then indicates that the Lanczos potential should also have fundamental significance in the way the 4-potential does in electrodynamics, and may be imbued with quantum aspects, but now in the presence of gravity.
Does it then mean that the Lanczos potential as a geometrical quantity opens a new gateway to the quantum world in the framework of a metric theory of gravity? There have already been some studies which seem to give an affirmative answer to the question.
Lanczos himself - the inventor of the potential tensor $L_{\mu\nu\sigma}$ - showed an intimate relation of his tensor to Dirac's equation describing an electron with spin \cite{Lanczos}. This is the reason why he named his tensor a `spintensor'.
In another study, Novello and Ridrigues \cite{N-R} have discovered a new interaction which can be thought of
as a short-range counterpart of gravitation, as weak interactions are the short-range
counterpart of electromagnetism. By using Lanczos tensor in Jordan's formulation of gravity, they have discovered a model in which gravity and electroweak interactions are described in a unique framework.

In the following, we discover some mores signatures of quantum physics that can be attributed to the Lanczos tensor.
Here, we want to emphasize that the exposition on the Lanczos theory, expounded in the preceding sections, does not consider the field equations of any particular theory of gravitation. Hence it holds in any metric theory of gravity formulated in a 4-dimensional pseudo-Riemannian spacetime.

\subsection{Singularity avoidance and Lanczos potential}

The existence of singularities in GR, where the classical spacetime curvature becomes infinitely large,  indicates a failure of the theory.
There have been claims that the singularities of GR can be resolved  by quantum effects.
Although a self-consistent theory of quantum gravity remains elusive, there is a general consensus that removal of classical gravitational singularities is not only a crucial conceptual test of any reasonable theory of  quantum gravity but also a prerequisite for it.

It has been shown that the classical cosmological singularities of various models can be avoided in quantum
cosmology \cite{Kiefer}.
Singularity avoidance also occurs in the framework of loop quantum cosmology, which provides a general scheme of singularity
removal that can be used for explicit scenarios \cite{Bojowald}.
We have already mentioned some signatures of quantum effects attributed to the Lanczos potential. Let us see if the potential  has any role in the avoidance of classical singularities.
As there does not exist any general agreement on the necessary criteria for quantum avoidance of
singularities, it would be sufficient to check if the potential itself avoids blowing up  at the singular points. 

The standard big-bang models are constructed by assuming the cosmological principle which leads to a homogeneous and isotropic spacetime represented by the R-W  line element (\ref{eq:RW}).
By solving Einstein's equation $R_{\mu\nu}-\frac{1}{2}g_{\mu\nu}R=-8\pi T_{\mu\nu}$ for (\ref{eq:RW}) and a perfect fluid $T_{\mu\nu}$, one gets the Friedmann models which show a singularity (big-bang) when the scale factor $S$ of the universe vanishes, signifying a state wherein the entire space shrinks to zero volume with the density going to infinity. 

Let us examine this situation in the accompanied Lanczos potentials derived for the line element (\ref{eq:RW}).
Since the curvature of the spacetime [given by the Ricci tensor in the case of the line element (\ref{eq:RW})] diverges at $S=0$ in general, one may expect the same fate for its Lanczos potential, particularly when it is the gravitational potential. Interestingly, all the Lanczos potentials obtained for this dynamic spacetime given by equations  (\ref{eq:L1-RW} - \ref{eq:L3-RW}, \ref{eq:Ln-RW}) are finite\footnote{The definitions (\ref{eq:L1-RW} - \ref{eq:L3-RW}, \ref{eq:Ln-RW}) of the Lanczos potential for the spacetime (\ref{eq:RW}) may not be exhaustive and hence cannot rule out the existence of any other value which blows up at $S=0$. (Such values, if exist, may however facilitate the possibilities of the existence of some non-vanishing minimum of $S$.) Nevertheless, the existence of the values of the tensor given by (\ref{eq:L1-RW} - \ref{eq:L3-RW}, \ref{eq:Ln-RW}), do substantiate our point of view considered in some particular gauges.} and well-defined at $S=0$ for a general $S=S(t)$ and for all values of the curvature parameter $k$. Let us recall that the only unknowns in the line element (\ref{eq:RW}) are $S(t)$ and $k$ which are determined and linked to the perfect fluid $T_{\mu\nu}$ by solving Einstein equation. Let us also recall that the coordinates $r,\theta$ appearing in equations  (\ref{eq:L1-RW}-\ref{eq:L3-RW}) are the comoving coordinates of the R-W spacetime (\ref{eq:RW}) and hence are independent of time. Thus the absence of any singularity in $L_{\mu\nu\sigma}$ given by these equations is beyond doubt.

On the one hand, the absence of a big bang-singularity in the Lanczos potential reassures that it is not a gravitational potential. On the other hand, it reveals a quantum signature of $L_{\mu\nu\sigma}$.
This may thus be helpful to develop an effective description of quantum gravity physics which
captures some quantum effects but is otherwise based on classical concepts.

\subsection{Gravitational analog of the Aharonov-Bohm effect}

At the classical level, a charged particle is considered to be influenced only by the electric and magnetic fields at the location
of the particle. At the quantum level however, the behaviour of a charged particle (confined to a region with vanishing electric and magnetic fields but non-vanishing 4-potential) is affected by the action of an external magnetic field from which the charged particle is excluded. This happens because the wave functions display a phase-shift due to non-vanishing potentials even in
regions where they give rise to no electrodynamic forces. This is the Aharonov-Bohm effect.

The electromagnetic analogy points out that the Lanczos potential should manifest an Aharonov-Bohm-like property in the case of gravity. 
In a metric theory of gravity, a genuine gravitational field is associated with a non-vanishing Riemann tensor.
The gravitational analog of the Aharonov-Bohm effect then suggests that the particles constrained to move in a region where the Riemann tensor vanishes, but not the Lanczos potential, may nonetheless exhibit physical effects which result from a non-vanishing curvature in a region from which the particles are excluded.

Numerous gravitational analogies of the Aharonov-Bohm effect have been studied in the past. Nevertheless, this has been done largely by considering the metric tensor $g_{\mu\nu}$  as the gravitational analog of  the electromagnetic potential  $A_\mu$ (see, for instance \cite{Vilenkin}).
However, we now know that this analogy is not quite correct and the true gravitational analog of the electromagnetic potential $A_\mu$  is the Lanczos tensor $L_{\alpha\beta\gamma}$.

A detailed study of the gravitational Aharonov-Bohm effect in terms of the Lanczos tensor is beyond the scope of the present article; a separate article itself dedicated to this topic is required.  Here we limit ourselves to showing the fulfillment of the minimum requirement expected from such an effect in the framework of a metric theory of gravity. That is, like its electromagnetic counterpart, the Lanczos tensor should be non-vanishing in the region where the particle is confined, viz. wherein the gravitational field vanishes and the spacetime becomes Minkowskian. This is indeed the case, as we shall see in the following - that the Lanczos tensor can very well be non-zero in the Minkowskian spacetime.

\subsubsection{Lanczos tensor for the Minkowskian spacetime}

Takeno has already shown that the  Minkowskian spacetime too admits, unexpectedly, non-trivial values for the Lanczos potential tensor \cite{Takeno}. Although he has derived differential equations for the Lanczos tensor for this spacetime, it has not been bracketed explicitly in closed form.
In order to derive it explicitly, let us consider the Minkowskian spacetime in the spherical polar coordinates:
\be
ds^2=dt^2- dr^2 - r^2d\theta^2 - r^2\sin^2\theta ~d\phi^2.\label{eq:Mink}
\ee
For this case, $C_{\mu\nu\sigma\rho}=0$ in equation (\ref{eq:Weyl-Lanczos2}) and the covariant derivatives reduce to the ordinary derivatives. With the aid of computational resources and taking guidance from the Takeno's differential equations for the Lanczos tensor derived for a spherically symmetric spacetime, we obtain the following with a single non-vanishing independent component as the Lanczos potential  for the line element (\ref{eq:Mink}):
\be
L_{rtt}= r f(t), ~~~~~  f(t)\equiv\text{an arbitrary function of} ~t, \label{eq:L1-Mink}
\ee
which is though not trace-free in its present form. A trace-free form can be obtained by using equation (\ref{eq:gaugeU}) giving
\be\label{eq:Lan-Mink}
\left.\begin{aligned}
&\bar{L}_{rtt}=\frac{2r}{3}f(t),\\
&\bar{L}_{r\theta\theta}=\frac{r^3}{3}f(t),\\
&\bar{L}_{r\phi\phi}=\frac{r^3}{3}f(t)\sin^2\theta.
 \end{aligned}
 \right\}
\ee
We also obtain another set of Lanczos potential for the line element (\ref{eq:Mink}) given by
\be
L_{trr}= g(r), ~~~~~  g(r)\equiv\text{an arbitrary function of} ~r, \label{eq:L2-Mink}
\ee
with a single non-vanishing independent component; and
\be\label{eq:Lan2-Mink}
\left.\begin{aligned}
&\bar{L}_{trr}=\frac{2}{3}g(r),\\
&\bar{L}_{t\theta\theta}=-\frac{r^2}{3}g(r),\\
&\bar{L}_{t\phi\phi}=-\frac{r^2}{3}g(r)\sin^2\theta.
 \end{aligned}
 \right\}
\ee
satisfying the gauge condition (\ref{eq:gauge}.a) of trace-freeness.
Clearly, a linear combination of two or more Lanczos potentials of this spacetime is again its Lanczos potential. 

Thus there indeed exist, by coincidence or providence, non-vanishing potentials supported by the Minkowski spacetime, justifying the existence of a gravitational analog of the Aharonov-Bohm effect. 
Assigning a `ground state'  potential field to the Minkowskian spacetime in the absence of any curvature, may appear puzzling and surprising at the first glance.
Nevertheless, the existence of this non-vanishing $L_{\mu\nu\sigma}$ with a vanishing $C_{\mu\nu\sigma\rho}$ is a reminiscent of and analogous to the non-vanishing electromagnetic potential $A_\mu$  outside a solenoid where $F_{\mu\nu}$ vanishes in the Chambers' experiment \cite{Chambers}.

The existence of the non-trivial potentials given by (\ref{eq:L1-Mink}-\ref{eq:Lan2-Mink}) for the flat Minkowskian spacetime with vanishing $C_{\mu\nu\sigma\rho}$, cannot be ignored by taking advantage of the arbitrariness in the functions $f(t), g(r)$  by assigning them to zero. This would be like forcing upon the theory a prejudiced interpretation, since a similar situation appears in the case of the R-W spacetime where one would not hesitate to accept a non-vanishing potential with $C_{\mu\nu\sigma\rho}=0$.

\subsection{Gravitational waves and Lanczos potential}

The fundamental observables of microscopic phenomena are described in terms of elementary particles and their collision.
In the parallel between electrodynamics and gravity underlies the fact that both interactions are mediated by massless particles - the photon and the graviton respectively. This is the reason why both classical theories look similar. 
Plane wave solutions of Maxwell's equations lead most naturally to an interpretation in terms of  the photon. Similarly, it is the wave solution in a gravitational theory, that is expected to lead to the concept of graviton.
Thus the theory of gravitational waves provides a crucial link between gravity and the microscopic frontier of physics.
Let us note that the interaction being mediated by virtual exchange of gravitons is also a prediction of a quantum theory of gravity.

It is already known (see, for example, \cite{IJGMMP,Wald}) that  a Killing vector field $A^\mu$ in a Ricci-flat  spacetime plays the role of the electromagnetic 4-potential and the source-free Maxwell equations, in Lorenz gauge ($A^\kappa_{~~;\kappa}=0$), reduce to the wave equation
\be
\nabla^\kappa \nabla_\kappa A_\mu \equiv g^{\alpha\beta} A_{\mu;\alpha;\beta} =0,\label{eq:Max}
\ee
where $\nabla^\kappa \nabla_\kappa$ is the curved-spacetime d'Alembertian operator.
In close correspondence with this, the Lanczos potential  too satisfies a homogeneous wave equation
\be
\nabla^\alpha\nabla_\alpha L_{\mu\nu\sigma} =0, \label{eq:waveL}
\ee
in Lanczos gauge, in any Ricci-flat spacetime \cite{Dolan-Kim}.
This simple and beautiful exact analytical solution has been paid the least attention while discussing the theory of gravitational waves in view of the recent observations of the gravitational waves emanated from the merger of binary black holes and neutron stars. Rather what is considered, in order to provide a theoretical explanation to these observations, is 
the numerical relativity simulations of Einstein's equation of GR. A simple wave equation on a par with equation (\ref{eq:waveL}), is a linearized approximation of Einstein's equation where the velocities are small and the gravitational fields are weak constrained by (\ref{eq:weak}). Einstein's equation then yields 
\be
\partial^\kappa \partial_\kappa \bar{h}_{\mu\nu}=0, ~~~~~ \bar{h}_{\mu\nu} \equiv h_{\mu\nu}-\eta_{\mu\nu}h/2, \label{eq:waveE}
\ee
in vacuum. Here $\partial^\kappa \partial_\kappa$ is the special-relativistic d'Alembertian operator.
However, the strongest gravitational-wave signals come
from highly compact systems with large velocities, i.e., from the processes where the linearization assumptions (\ref{eq:weak}) do not apply. Thus 
equation (\ref{eq:waveE}) is not competent to explain accurately the gravitational-wave emission from the violent processes like the stellar core collapse and the mergers of black holes or neutron stars. But equation (\ref{eq:waveL}) consistently serves the purpose, which holds in the most general case (in a Ricci-flat spacetime) in any metric theory of gravity formulated in a 4-dimensional pseudo-Riemannian spacetime.
In a general case where $R_{\mu\nu}$  is not necessarily vanishing, the corresponding wave equation is obtained as
\be
\nabla^\alpha\nabla_\alpha L_{\mu\nu\sigma} =\psi(R_{\mu\nu},  L_{\mu\nu\sigma}), \label{eq:waveL-general}
\ee
where the function $\psi$ can be calculated in terms of the matter energy-momentum tensor by assuming a field equation, for instance Einstein's equation in the case of GR. 

Thus the Lanczos potential theory provides an outline of a classical gravitational field whose quantum description would be a massless spin-2 field propagating at the speed of light. 
Let us recall that in order to understand gravity on the same footing as the other interactions, one has to consider
it as a spin-2 gauge theory.

\section{Summary and Conclusion}

Einstein's enlightening insight - the local equivalence of gravitation and inertia - paved way for the geometrization of gravitation in the framework of a pseudo Riemannian spacetime.
By considering the well-noted correspondence between gravitation and electrodynamics, we have developed another insight that the geometry of the spacetime is endowed with at least two fundamental geometric structures. First, the Riemann-Christoffel curvature tensor which is the nodal point for the unfolding of gravity in any metric theory. Second, a rank-three tensor discovered by Lanczos which is enriched with extraordinary scientific and philosophical value, but has nevertheless gone largely unnoticed by mainstream relativists and cosmologists. The Lanczos tensor, which appears as the potential for the Weyl tensor, emerges as an inherent structural element of any metric theory of gravity formulated in a 4-dimensional pseudo Riemannian spacetime, without considering the field equations of any particular theory.

By deriving expressions for the Lanczos tensor in some particularly chosen spacetimes, we have attempted to find its physical meaning and an adequate interpretation. It appears that the tensor does not represent a relativistic formulation of the potential of the gravitational field, despite being assigned to the potential of the Weyl tensor which shares a major part of the curvature of spacetime.
Rather, it is impregnated with signatures of quantum physics and opens up a new gateway to the quantum world in the framework of a metric theory of gravity. This is ascertained by various evidences which open up a novel vision in a geometric embodiment of gravity.

It appears that the consequences of the geometrization of gravitation go beyond what we know today and rich prospects stand open for investigation by considering the lead of the Lanczos potential tensor.

\bigskip
\noindent
{\bf Acknowledgements:}
The author gratefully acknowledges useful discussions on various topics of
this work with Jayant V. Narlikar and Sanjeev V. Dhurandhar.
He would like to thank IUCAA for the hospitality, where this work was initiated during a visit. 
He also acknowledges the use of the computer algebra package ``GRTensor'' in many calculations.

\end{document}